\documentclass[twocolumn,pra,aps,showpacs,superscriptaddress,10pt]{revtex4-1}  
\usepackage{amsmath,amssymb,graphicx}
\usepackage[colorlinks, citecolor=blue,linkcolor=blue]{hyperref}
\usepackage{hyperref}
\usepackage{epsfig}
\begin{document}
\title{Efficient tunable switch from slow light to fast light in quantum opto-electromechanical system}
\author{M. Javed Akram}\email{javed\_quaidian@yahoo.com}
\affiliation{Department of Electronics, Quaid-i-Azam University, 45320 \ Islamabad, Pakistan.}
\author{Khalid Naseer}
\affiliation{Department of Physics, University of Sargodha, Sargodha, Pakistan}
\author{Farhan Saif}
\affiliation{Department of Electronics, Quaid-i-Azam University, 45320 \ Islamabad, Pakistan.}
\date{\today}
\begin{abstract}
The control of slow and fast light propagation, in the probe transmission in a single experiment, is a challenging task. This type of control can only be achieved through highly nonlinear interactions and additional interfering pathway(s), which is therefore seldom reported. Here, we devise a scheme in which slow light, and a tunable switch from slow light to fast light can be achieved in the probe transmission based on a hybrid setup, which is composed of an optical cavity with two charged nano mechanical resonators (MRs). The two MRs are electrostatically coupled via tunable Coulomb coupling strength ($g_c$) making a quantum opto-electromechanical system (QOEMS). The parameter $g_c$ that couples the two MRs can be switched on and off by controlling the bias voltages on the MRs, and acts as a tunable switch that allows the propagation of transmitted probe field as slow light ($g_c\neq0$) or fast light ($g_c=0$). In our scheme, the magnitude of delay and pulse advancement can be controlled by tuning the Coulomb interaction and power of the pump field. Furthermore, we show that slow light regime in our model is astonishingly robust to the cavity decay rate. In comparison with previous schemes, our scheme has clear advantages that empowers the state-of-the-art photonic industry as well as reflects the strength of emerging hybrid technologies.
\end{abstract}
\maketitle
\section{Introduction}
Electromagnetically induced transparency (EIT) \cite{1,2,3,4,5,6,7}, which is a quantum interference phenomenon that takes place when two laser fields excite resonantly and two different transitions share a common state, provides counter-intuitive ways that how a strong laser field can dress a dielectric medium which leads to the modification of the absorptive and dispersive characteristics of the medium \cite{8}. This alters the group index of the medium such that group velocity of a weak probe laser injected in the dressed medium can be modified in a controllable fashion, which engenders slow and fast (a.k.a. superluminal) light effects \cite{8,9,10,11,12,13}.

Various techniques have been developed to realize slow and fast light in atomic vapors and solid-state media \cite{8,11}. Techniques related to photon-echo spectroscopy, such as atomic frequency combs \cite{14} and controllable reversible inhomogeneous broadening \cite{15}, have been used to efficiently achieve quantum memories via quantum storage and optical buffering. On the other hand, the rapid dephasing rates and inhomogeneous broadening of solid-state electronic resonances, have suggested a plethora of methods and techniques to control the propagation of light. For instance, the techniques based on stimulated Brillouin scattering (SBS) in fibers \cite{16} and coherent population oscillations (CPO) \cite{17} have been used to significantly delay classical light fields. In chip-scale photonics, dynamically tunable arrays of cavities in EIT configuration, are an intriguing analogue to ensembles of atoms and provide a route to slowing and even stopping light pulses all-optically \cite{18,19}.

Recently, the field of cavity optomechanics has cemented present-day photonic technology in view of its promising applications from microscopic to macroscopic domains \cite{20,21,22,23}. In optomechanical resonators, an optical mode couples to mechanical vibrations via radiation pressure induced by circulating optical fields \cite{20}. Generally, the combination of additional matter-like systems (for example, mechanical membranes \cite{24,25,27,26}, single multi-level atoms \cite{28,29,30,31,32,31} and Bose Einstein Condensate/Fermions \cite{33,34,35}) with these opto-mechanical resonators, leads to the emergence of hybrid quantum systems \cite{22,36,37}, which serve as basic building blocks from quantum state-engineering to the quantum communication networks \cite{38,39,40}. Due to the ubiquitous nature of the mechanical motion, such resonators couple with many kinds of quantum devices, from atomic systems to the solid-state electronic circuits \cite{32,Tian}. An excellent example of such a system is quantum electromechanical system (QEMS) \cite{46,47,Tian}. QEMS is a device \cite{46,47,Tian,52,522} where, in the observable behavior, the quantum nature of either the electronic or mechanical degrees of freedom becomes important. Such a system was earlier suggested by Wineland {\it et al.} \cite{44,45}, and later by Zoller \& Tian \cite{46} and Milburn and co-workers \cite{47}. In QEMS, the two charged cantilevers are coupled electrostatically via Coulomb interaction \cite{47,44,Bachtold}. In cavity optomechanical and electromechanical systems, the resolved-sideband regime can now be reached with the mechanical frequency far surpassing the cavity bandwidth \cite{53,54,55,56,57,Bennett}. Moreover, mechanically-induced transparency due to the destructive interference of the probe field with the anti-Stokes field generated by the pump field \cite{4,6} and normal-mode splitting in the strong coupling limit \cite{58,59} have been reported.

Our report is based on combining these electromechanically coupled charged cantilevers with optomechanics, that bridges the traditional quantum optics based OMS with electrostatics based QEMS, providing us a new hybrid testbed, viz. quantum opto-electromechanical system (QOEMS) \cite{Tian,48,50,51,Wang,Tsang,Terry,Regal}. In analogy with multi-level atomic systems, hybrid optomechanics empowers us with a coherent control to the engineered couplings that give rise to Fano-like interferences \cite{5,6,7,Clerk} and even double EIT phenomenon \cite{7,60,61}. Here, using the hybrid QOEMS, we demonstrate a scheme that provides us a tunable switch from slow light to fast light in the probe transmission.
Unlike previous schemes \cite{63,64,65,66}, our scheme owns some favorable features: {\it (i)} Two transmission windows appear in our model, due to the additional MR, that causes the high dispersion and leads to the switching mechanism from slow light to fast light in the probe transmission. {\it (ii)} Slow light propagation can be observed when both the coupling parameters contribute together, which can be adjusted by controlling the Coulomb coupling strength and the pump power. {\it (iii)} Slow light regime in our model is astonishingly robust to the cavity decay rate. {\it (iv)} A tunable switch from slow light to fast light can be obtained by turning off the Coulomb coupling strength as $g_c=0$, which can be adjusted by the pump power.  {\it (v)} Both slow and fast light effects can be observed in the probe transmission in a single experiment.
\section{The Model Formulation}\label{sec2}
A schematic representation of our proposed model is illustrated in Fig.~\ref{setup}. We consider a realistic optomechanical system, composed of a high $Q$ Fabry-P\'{e}rot cavity of length $L$, with a fixed mirror and a movable nano mechanical resonator (MR). Specifically, the mechanical resonator consists of two charged MRs, where the $MR_{1}$ of the optomechanical cavity couples to the cavity field via radiation pressure, as well as electrostatically coupled to another movable oscillator outside the cavity (i.e., MR$_{2}$) through the tunable Coulomb coupling strength, viz. $g_{c}$. Such an eletromechanical subsystem in our QOEMS was earlier suggested by Wineland {\it et al.} \cite{44,45}, and later by Milburn and co-workers \cite{47}, and Zoller \& Tian \cite{46} where an ion is trapped between two parallel suspended electrodes represented by MRs (or nanowires or nanotubes). This coupling can be switched on and off using an external bias voltage at an electrode on the oscillators, where MR$_{1}$ is charged by the bias gate voltage $V_{1}$ and subject to the Coulomb force due to another charged MR$_{2}$ with the bias gate voltage $-V_{2}$. Thus, the system has usual optomechanical coupling ($g_{o}$), and the additional tunable Coulomb coupling ($g_c$), forming a hybrid quantum eletro-optomechanical system (QOEMS). The optomechanical cavity is simultaneously driven by an intense pump field of frequency $\omega_l$, and a weak probe field of frequency $\omega_p$, along the cavity axis. Hence, under the rotating reference frame at the frequency $\omega_l$ of the strong driving field, the total Hamiltonian of the system can be written as:
\begin{figure}[t]
\centering
\includegraphics[width=0.5\textwidth]{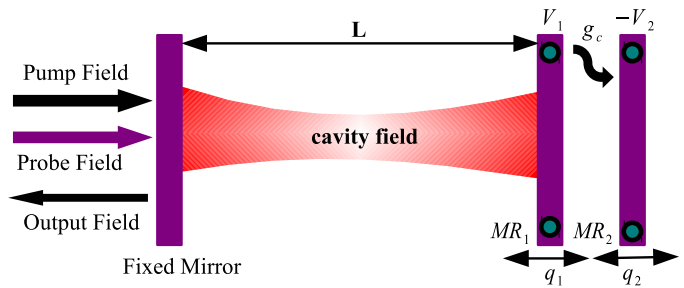}
\caption{The schematic of the opto-electromechanical system. A high $Q$ Fabry-P\'{e}rot cavity of length $L$ consists of a fixed mirror and a movable mirror MR$_{1}$ with small displacement $q_1$. MR$_{1}$ is charged by the bias gate voltage $V_{1}$, and subject to the Coulomb force due to another charged MR$_{2}$ with the bias gate voltage $-V_{2}$. The parameter $g_c$ represents the Coulomb coupling strength between the two mechanical oscillators.}\label{setup}
\end{figure}
\begin{eqnarray}
&&H_T=[\frac{p_1^2}{2m_1}+\frac{1}{2}m_1\omega_1^2q_1^2]
+[\frac{p_2^2}{2m_2}+\frac{1}{2}m_2\omega_2^2q_2^2] \notag \\ &&+ \hbar\Delta_cc^\dag c -\hbar g_o c^\dag cq_1
+\hbar g_c q_1q_2  +i\hbar\Omega_l(c^\dag -c) \notag \\
&&+ i\hbar(\varepsilon_p e^{-i\delta t}c^\dagger -\varepsilon_p^* e^{i\delta t}c), \label{Ham}
\end{eqnarray}
where $\Delta_c=\omega_c-\omega_l$ and $\delta=\omega_p-\omega_l$  are the detunings of the cavity field frequency and the probe field frequency respectively, with strong pump field frequency $\omega_l$.
In Eq.~(\ref{Ham}), first two terms describe the free Hamiltonian of the moving mirrors (MR$_{1}$ and MR$_{2}$). The operators, $q_i$ and $p_i$, represent the position and momentum operators of both the mirrors respectively, vibration frequency $\omega_{i}$, and effective mass $m_{i}$. Third term represents single-mode of cavity field with frequency $\omega_{c}$ and annihilation (creation) operator $c~(c^{\dag})$. The fourth term expresses optomechanical coupling between MR$_{1}$ and the cavity field with the coupling strength $g_o=\frac{\omega_{c}}{L}\sqrt{\hbar/m\omega_1}$. The fifth term represents the Coulomb coupling between the charged MRs with the Coulomb coupling strength $g_c =\frac{C_{1}V_{1}C_{2}V_{2}}{2\pi \hbar \epsilon _{0}x_{0}^{3}}$ \cite{47, 60, 46}; where MR$_{1}$ and MR$_{2}$ take the charges $C_{1}V_{1}$ and $-C_{2}V_{2}$, with $C_{1}(C_{2})$ and $V_{1}(-V_{2})$ being the capacitance and the voltage on the mirrors respectively, and $x_0$ is the equilibrium separation between the two oscillators. Finally, the last two terms correspond to the classical light fields (pump and probe fields) with frequencies $\omega_l$ and $\omega_p$, respectively. Furthermore, $\Omega_l$ and $\varepsilon_p$ are related to the laser power $P$ by $|\Omega_l|=\sqrt{2\kappa P_{l}/ \hbar \omega_l}$, and $\varepsilon_p=\sqrt{2\kappa P_{p}/ \hbar \omega_p}$.

Since here we deal with the mean response of the coupled system to the probe field, we identify the expectation values of all the operators. By taking the corresponding dissipation and fluctuation terms into account, and using the mean field approximation \cite{7}, i.e. $\langle qc\rangle =\langle q\rangle \langle c\rangle $, the mean value equations are given by,
\begin{eqnarray}\label{mve}
&&\langle \dot{q_{1}}\rangle =\frac{\langle p_{1}\rangle }{m_{1}},  \notag
~~~ \langle \dot{q_{2}}\rangle =\frac{\langle p_{2}\rangle }{m_{2}},  \notag \\
&&\langle \dot{p_{1}}\rangle =-m_{1}\omega _{1}^{2}\langle
q_{1}\rangle -\hbar g_{c} \langle q_{2}\rangle +\hbar g_o\langle
c^{\dag }\rangle \langle
c\rangle -\gamma _{1}\langle p_{1}\rangle ,  \notag \\
&&\langle \dot{p_{2}}\rangle =-m_{2}\omega _{2}^{2}\langle
q_{2}\rangle -\hbar g_{c} \langle q_{1}\rangle -\gamma _{2}\langle
p_{2}\rangle,
\notag \\
&&\langle \dot{c}\rangle =-(\kappa +i\Delta _{c})\langle c\rangle + ig_o\langle q_{1}\rangle \langle c\rangle +\Omega_{l}+\varepsilon_{p}e^{-i\delta t}.
\end{eqnarray}

Here, $\gamma _{1}$ and $\gamma _{2}$ are the decay rates associated with MR$
_{1}$ and MR$_{2}$, respectively. As we encounter the expectation values of all the operators in Eq.~(\ref{mve}), therefore, we drop the quantum Brownian noise and input vacuum noise terms which average to zero. In order to acquire the steady-state solutions of the above equations, we make the ansatz~\cite{7,6}: $
\langle h \rangle = h_s +  h_- e^{-i\delta t} + h_+ e^{i\delta t}$,
where, $h_s$ denotes any of the steady-state solutions $c_s$, $q_{is}$ and $p_{is}$. Moreover, $h_\pm$ are of the same order as $\varepsilon_p$. In the case of $h_{s}\gg h_{\pm }$, Eq. (\ref{mve}) can be solved by treating
$h_{\pm }$ as perturbations. Upon substituting the ansatz into Eqs.~(\ref{mve}) and upon working to the lowest order in $\varepsilon_p$, we obtain the following steady-state solutions:
\begin{eqnarray}\label{mean}
&&\notag q_{1s}=\frac{\hbar g_o|c_{s}|^{2}}{m_{1}\omega _{1}^{2}-\frac{\hbar
^{2}g_c ^{2}}{m_{2}\omega _{2}^{2}}}, ~~~
q_{2s}=\frac{-\hbar g_c q_{1s}}{m_{2}\omega _{2}^{2}}, ~~~    \\
&&c_{s}=\frac{\Omega_{l}}{\kappa+i\Delta}, \\
&&c_{-}=\frac{[\kappa -i(\Delta +\delta )][(\delta ^{2}-\omega
_{1}^{2}+i\delta \gamma _{1})-\alpha] - 2i\omega _{1}\beta }{[\Delta
^{2}-(\delta +i\kappa )^{2}][(\delta
^{2}-\omega_{1}^{2}+i\delta\gamma_{1}) - \alpha]+4\Delta \omega_{1}\beta }. \nonumber
\end{eqnarray}
Here, $\alpha=\frac{ \hbar^2g_c^{2}}{m_1m_2(\delta
^{2}-\omega_{2}^{2}+i\delta\gamma_{2})}$, $\beta =\frac{\hbar g_o^{2}|c_{s}|^{2}}{2m\omega_{1}}$, and $\Delta=\Delta_c - gq_{1s}$ is the effective detuning. In order to investigate the optical property of the output field, we use the standard input-output relation \cite{4}: $c_{out}(t) = c_{in}(t) - \sqrt{2\kappa}c(t)$, where $c_{in}$ and $c_{out}$ are the input and output operators, respectively. By using the above input-output relation and the ansatz for $\langle c(t) \rangle$, we can obtain the expectation value of the output field as \cite{13,6},
\begin{eqnarray}
\langle c_{out}(t) \rangle &=& (\Omega_l-\sqrt{2\kappa} c_s) + (\varepsilon_p - \sqrt{2\kappa} c_-) e^{-i\delta t} \notag \\
&-& \sqrt{2\kappa} c_+ e^{i\delta t}.
\end{eqnarray}

Now, the transmission of the probe field, which is ratio of the returned probe field from the coupling system divided by the sent probe field \cite{6,4}, can be acquired as
\begin{equation}
t_p(\omega_p) = \frac{\varepsilon_p - \sqrt{2\kappa} c_-}{\varepsilon_{p}} = 1- \frac{\sqrt{2\kappa} c_-}{\varepsilon_p}.
\end{equation}
Importantly, for an optomechanical system, the presence of the pump laser not only induce a strong modification of the transmission of the probe field, but also leads at the same time to a rapid variation of the phase of the transmitted probe field across the transmission window. The rapid phase dispersion \cite{6}, that is, $\phi_t(\omega_p) = arg[t_p(\omega_p)]$, can
lead to significant group delays in analogy to that we achieve with EIT in atomic \cite{69,70} and in solid-state media \cite{13,6}. The transmission group delay \cite{4,6} is given as
\begin{equation}
\tau_g = \dfrac{d\phi_t(\omega_p)}{d\omega_p}=\dfrac{d\{arg[t_p(\omega_p)]\}}{d\omega_p}, \label{phase}
\end{equation}
where, $\tau_g > 0$ and $\tau_g < 0$ respectively, correspond to slow and fast light propagation \cite{8,6,13}. In the following section, we show that two transmission windows occur due to additional mechanical oscillator in the hybrid opto-electromechanical system. The parameter $g_c$ serves as tunable switch that causes the normal dispersion (slow light) when it is present, and yields anomalous dispersion (fast light) when it is zero (or off). Consequently, we expect slow and fast light regimes in the probe transmission in hybrid QOEMS.
\section{Results and Discussions}\label{sec3}
In order to quantify the slow and fast light effects of the transmitted probe field, in our numerical simulations, we choose experimentally realizable parametric values of the optomechanical system. We use the parameters from recent experiment \cite{57}: $L=$25 mm, $\lambda=$1064 nm, $\omega_i/ 2\pi = 947$ KHz ($i=1,2$), $Q_{i}=\frac{\omega_i}{\gamma_i}=6700$, $m_i=145$ ng, $\kappa/2\pi=215$ KHz, $P_l=6$ $\mu$W, and $g_c/2\pi=8 \times 10^{6}$ Hz/m$^{2}$ \cite{46}. Note that $\omega_1 > \kappa$, therefore, the system operates in the resolved-sideband regime \cite{67,6}. In the following subsections, we explain the occurrence of slow and fast light regimes in the QOEMS, for $g_c \neq 0$ and $g_c=0$, respectively.
\begin{figure}[htb]
\includegraphics[width=.48\textwidth]{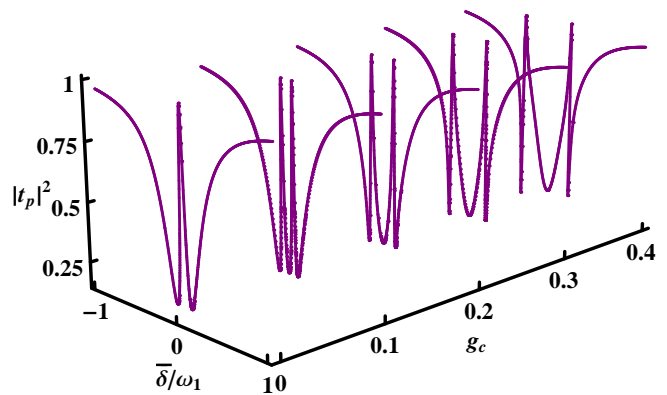}
\caption{Plot of the tranmission $|t_p|^2$ versus the normalized probe detuning as a function of the Coulomb interaction $g_c$ (in units of $2\pi \times g_c$ MHz/m${^2}$). The experimental parameters are: $L=25$ mm,  $\lambda=$1064 nm, $\omega_{1,2}/2\pi=947$ KHz, $Q_{1,2}=\frac{\omega_{1,2}}{\gamma_{1,2}}=6700$, $m_{1,2}=145$ ng, $\kappa/2\pi=215$ KHz, and $\Omega_l/2\pi=2$ MHz \cite{57}. For $g_c=0$, single transmission window appears, and two trasmission windows occur once the Coulomb coupling is present.}\label{tr1}
\end{figure}
\subsection{Slow light regime}
We consider Coulomb interaction ($g_{c}$) between the two oscillators together with optomechanical coupling $g_{o}$, in the QOEMS, to control the propagation of the probe field. In Fig.~\ref{tr1}, we plot the transmission spectrum of the probe field as a function of normalized probe detuning $\overline{\delta}/\omega_{1}$ for different values of Coulomb coupling $g_{c}$. Here, $\overline{\delta}=\delta-\omega_1$ is the probe detuning from the line center. In the absence of the Coulomb coupling i.e. $g_c=0$, in Fig.~\ref{tr1}, we see that a single peak in the transmission spectrum appears, at $\overline{\delta} = 0$ ($\overline{\delta} = \omega_1$) in agreement with a previous report for a single MR \cite{4}.

However, the characteristics of the transmitted probe field changes in the presence of the Coulomb coupling between the two MRs and remarkably two windows occur in the probe transmission spectrum with different centers. Figure.~\ref{tr1} illustrates that increasing the value of the Coulomb coupling broadens the central window, and the two central peaks in the transmission spectrum are further and further apart. Analytically, we note that when there is no Coulomb coupling, $g_c=0$ (i.e. $\alpha=0$) between the two MRs, see Eq.~(\ref{mean}), which reduces to Eq.~(5) in Ref.~\cite{67}. However, different from the output field in Ref.~\cite{67} which involves a single center frequency for the single transparency window, there are two centers with different frequencies in our model due to the Coulomb interaction. Consequently, under the action of the radiation pressure and the probe field, two transparency windows appear in the transmission profiles. This effect is analogous to the phenomenon of double EIT, that has been extensively reported in various hybrid systems, e.g. in plasmonic nanocavity \cite{68}, hybrid optomechanics \cite{7} and optomechanical ring cavity \cite{62}.

\begin{figure}[t]
\includegraphics[width=0.45\textwidth]{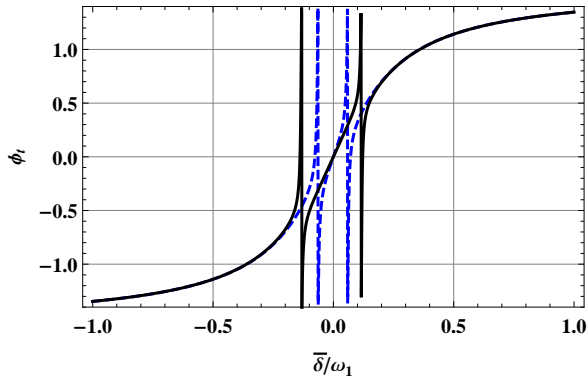}
\caption{The phase, $\phi_t$, of the probe field as a function of normalized probe detuning, $\overline{\delta}/\omega_1$, subject to the Coulomb interaction, for $g_c/2\pi=0.1$ MHz/m$^2$ (dashed-blue curve) and $g_c/2\pi=0.2$ MHz/m$^2$ (solid-black curve). All parameters are the same as in Fig.~\ref{tr1}.} \label{ph1}
\end{figure}
\begin{figure}[t]
\includegraphics[width=0.5\textwidth]{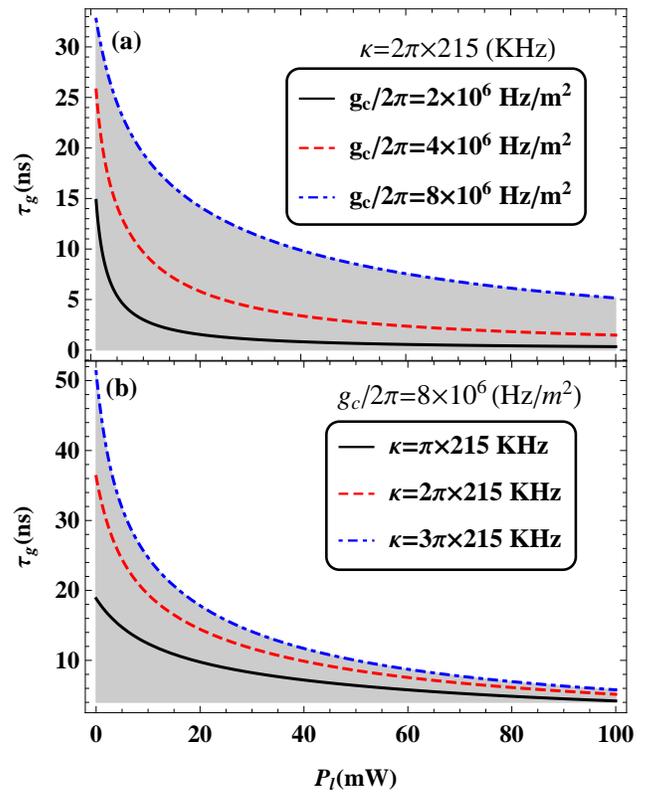}
\caption{(a) Plot of the group delay $\tau_g$ as a function of the pump power $P_l$ for $\kappa=2\pi \times 215$ KHz, and $g_c/2\pi=2,4,8$ MHz/m$^2$ (from bottom to top). (b) Plot of $\tau_g$ as a function of the pump power $P_l$, for $g_c/2\pi=8$ MHz/m$^2$, and $\kappa=\pi \times 215$, $2\pi \times 215$, $3\pi \times 215$ KHz (from bottom to top). All parameters are the same as in Fig.~\ref{tr1}.} \label{delay1}
\end{figure}
In Fig.~\ref{ph1}, we present the phase of the probe field versus the normalized probe detuning $\overline{\delta}/\omega_{1}$ for $g_c/2\pi=0.1,0.2$ MHz/m$^2$ (solid curve, dashed curve). Due to the presence of two MRs, we observe a high positive dispersion indicating the occurrence of slow light in the probe transmission. In the control of slow light, it is always advantageous to obtain the high dispersion as it greatly alters the group delay of the pulse injected in the transparent medium \cite{8,6}. Here, we show that by exploiting the electromechanical coupling $g_c$ in the hybrid QOEMS, high dispersion is acquired.

In order to examine the quantitative features of the probe propagation with subject to the electromechanical coupling $g_c$, we present the group delay $\tau_g$ versus the pump power $P_l$ in Fig.~\ref{delay1}. A positive group delay indicates a slow light in the transmitted probe field in the presence of $g_c$ as expected. We note that: {\it (i)} slow light occurs when the Coulomb interaction dominates the optomechanical coupling, and the magnitude of the delay can be increased with increase in $g_c$, as shown in Fig.~\ref{delay1}(a). Here $\kappa$ is fixed, and $g_c$ takes different values.
{\it (ii)} In Fig.~\ref{delay1}(b), group delay $\tau_g$ is shown for $g_c/2\pi=8$ MHz/m$^2$, and for the different values of $\kappa$. Surprisingly, we see that with increase in $\kappa$, the magnitude of delay increases, which reflects that our slow light regime is robust against the cavity decay rate, which is seldom reported before. This directly confirms the occurrence of slow light regime mentioned in {\it (i)}, which shall become clear in the later discussion where we obtain the superluminal behavior of the probe field by keeping the parameter, $g_c$, as switched off. At grass-root level, the underlying mechanism behind the robustness of slow light confirms the basis of cavity optomechanics and theory of radiation pressure. This can be understood as follows: Under the constant Coulomb interaction and the constant drive (pump field), the equilibrium position follows the strain of the MR. With the increase in $\kappa$, the radiation pressure on MR$_{1}$ decreases, while both the MRs will acquire large displacement to provide large strain in order to compensate the reduced radiation pressure \cite{20,Tian}. The transmission spectrum becomes narrower for the larger displacement of the MRs. In this way, slow light regime in QOEMS is robust against the cavity decay rate $\kappa$.
\subsection{Fast light regime}
\begin{figure}[ht]
\includegraphics[width=0.45\textwidth]{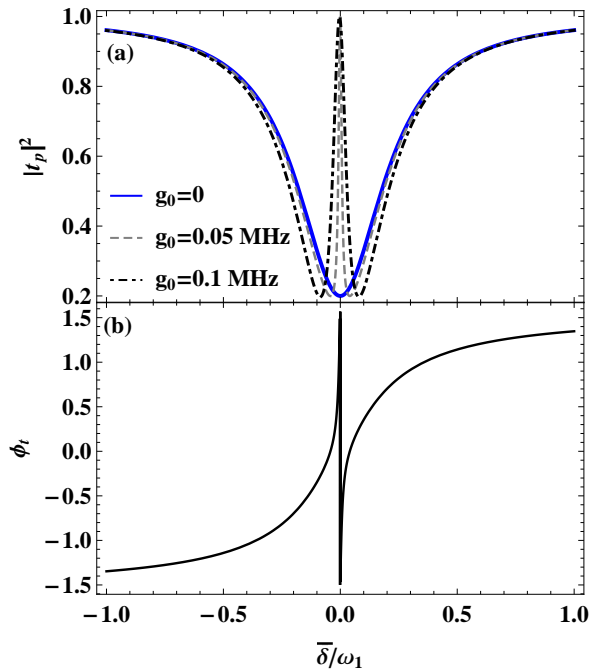}
\caption{(a) The transmission $|t_p|^2$ of the probe field as a function of normalized probe detuning $\overline{\delta}/\omega_1$ in the absence of the Coulomb interaction for $g_{o}=0,0.05,0.1$ MHz, respectively. (b) The phase $\phi_t$ of the probe field versus the normalized probe detuning for $g/2\pi=0.1$ MHz in the absence of the Coulomb interaction, i.e. $g_{c} \simeq 0$. All parameters are the same as in Fig.~\ref{tr1}.} \label{tr2}
\end{figure}

Here, we consider the case when Coulomb coupling is switched off or very small, i.e. $g_c\simeq 0$. This reduces the hybrid QOEMS to a single-ended OMS. For the present case, in Fig.~\ref{tr2}, we show the transmission $|t_p|^2$ of the probe field as a function of normalized probe detuning ($\overline{\delta}/\omega_1$), for different values of optomechanical coupling $g_o$. We see that, in the absence of pump laser, i.e. $g_{o}=0$, a standard Lorentzian curve \cite{4,13} appears (thick-solid curve) as shown in Fig.~\ref{tr2}.
\begin{figure}[ht]
\includegraphics[width=0.48\textwidth]{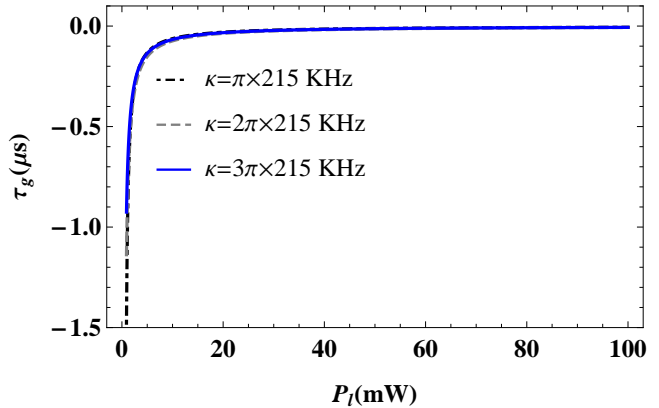}
\caption{Plot of the the group delay $\tau_g$ as a function of the pump power $P_l$, for different values of cavity decay rate $\kappa$. All parameters are the same as in Fig.~\ref{tr2}.} \label{ph2}
\end{figure}
However, in the presence of intense pump laser, the transmission spectrum of the probe field shows a prominent transparency window at the resonant region, viz., $\overline{\delta} \sim \omega_1$. Form Fig.~\ref{tr2}(a), it can be seen that transparency window broadens and becomes more transparent by increasing the optomechanical coupling, in agreement with previous reports \cite{13,4,6,63}. Nevertheless, it reflects that our approach is an extension of previously reported single-ended cavities, here our motivation is to acquire a tunable switch that permits us to explicitly control both slow light and fast light propagation in the probe transmission.

In Fig.~\ref{tr2}(b), we plot the phase as a function of normalized probe detuning. Note that, the phase of the probe field suffers a steep negative dispersion around $\overline{\delta}=0$, as shown in Fig.~\ref{tr2}(b). This rapid phase change leads to the change in group delay of the probe field such that fast light effect takes place. Figure.~\ref{ph2} displays the group delay for the different values of cavity decay rate $\kappa$. The group delay is negative, which reveals that we observe the fast light propagation (pulse advancement) in the transmitted probe field, when $g_c$ is switched off. From figure~\ref{ph2}, it can be seen that for decrease in $\kappa$, magnitude of pulse advancement increases. Thus, longer the lifetime of the resonator, more obvious to see the superluminal behavior.

We emphasis that we observed both the phenomena, slow and fast light behavior of the probe field through the transmission. However previously it is reported that the slow light occurs in the probe transmission and fast light occurs in the reflected field \cite{6,63,4}. Thus, we report a tunable switch based on a transition from slow light to fast light as a function of $g_c$, and the robustness of slow light against the dissipation. This clarifies the advantage of our scheme in contrast to the earlier reports \cite{13,63,65,64,66}.
\section{Conclusion}\label{sec4}
The following conclusions apply to our reported work: (i) By utilizing the hybridization of optomechanical and electromechanical systems, we report that both slow and fast light behaviors can be achieved in a single hybrid experimental setup. (ii) In contrast to the previous schemes, two tunable transmission windows occur in our model due to the additional MR, that provides us a tunable switching mechanism from slow light to fast light. (iii) We observed slow light in the transmitted probe field in the presence of Coulomb coupling strength, and witnessed that slow light regime in our model is robust to the cavity decay rate. (iv) A tunable switch from slow light to fast light can be obtained in our model by adjusting the Coulomb coupling strength as $g_{c}=0$. (v) Both pulse delay and advancement can be observed in the probe transmission in a single experiment. Slow and fast light characteristics demonstrate great potential in many applications, including integrated quantum optomechanical memory, designing novel quantum information processing gates, classical signal processing, optical buffers for telecommunication systems, delay lines and interferometry.


\begin{thebibliography}{99}
\bibitem{1} S. E. Harris, Phys. Today {\bf 50}, 36 (1997).

\bibitem{2} M. Fleischhauer, A. Imamoglu, and J. P. Marangos, Rev. Mod. Phys. {\bf 77}, 633 (2005).

\bibitem{3} M. D. Lukin and A. Imamoglu, Nature (London) {\bf 413},
273 (2001).

\bibitem{4} S. Weis {\it et al}., Science {\bf 330}, 1520 (2010).

\bibitem{5} G. S. Agarwal and S. Huang, Phys. Rev. A {\bf 81}, 041803(R) (2010).

\bibitem{6} A. H. Safavi-Naeini, T. P. Alegre, J. Chan, M. Eichenfield,
M. Winger, Q. Lin, J. T. Hill, D. E. Chang, and O. Painter,
Nature (London) {\bf 472}, 69 (2011).

\bibitem{7} M. J. Akram, F. Ghafoor, and F. Saif, J. Phys. B: At. Mol. Opt. Phys. {\bf 48}, 065502 (2015).

\bibitem{8} P. W. Milonni, Fast Light, Slow Light and Left-handed Light ({\it Taylor \& Francis}, 2010).

\bibitem{9} L. V. Hau, S. E. Harris, Z. Dutton, and C. H. Behroozi, Nature
(London) {\bf 397}, 594 (1999).

\bibitem{10} M. S. Bigelow, N. N. Lepeshkin, and R. W. Boyd, Science {\bf 301}, 200 (2003).

\bibitem{11} R. W. Boyd and D. J. Gauthier, Science {\bf 326}, 1074 (2009).

\bibitem{12} M. O. Scully and M. S. Zubairy, Science, {\bf 301}, 181 (2003).

\bibitem{13} M. J. Akram, and F. Saif, arXiv:1501.06062 [quant-ph] (2015).

\bibitem{14} H. de Riedmatten, M. Afzelius, M. U. Staudt, C. Simon, and N. Gisin, Nature (London) {\bf 456}, 773 (2008).

\bibitem{15} M. Afzelius, C. Simon, H. de Riedmatten, and N. Gisin, Phys. Rev. A {\bf 79}, 052329 (2009).

\bibitem{16} L. Th\'evenaz, Nature Photon. {\bf 2}, 474 (2008).

\bibitem{17} M. S. Bigelow, N. N. Lepeshkin, and R. W. Boyd, Science {\bf 301}, 200 (2003).

\bibitem{18} M. F. Yanik, W. Suh, Z. Wang, and S. Fan, Phys. Rev. Lett. { \bf 93}, 233903 (2004). A. Peng, M. Johnsson, W. P. Bowen, P. K. Lam, H.-A. Bachor and J. J. Hope, Phys. Rev. A {\bf 71}, 033809 (2005); {\it ibid.} {\bf 74}, 059902(E) (2006).

\bibitem{19} D. E. Chang, A. H. Safavi-Naeini, M. Hafezi, and O. Painter,
New J. Phys {\bf 13}, 023003 (2011).

\bibitem{20} M. Aspelmeyer, T. J. Kippenberg, F. Marquardt, Rev. Mod. Phys. {\bf 86}, 1391 (2014).

\bibitem{21} P. Meystre, Ann. Phys. (Berlin) {\bf 525}, 215 (2013).

\bibitem{22} B. Rogers, N. Lo Gullo, G. De Chiara, G. M. Palma, and M. Paternostro, Quantum Meas. Quantum Metrol. {\bf 2}, 11-43 (2014).

\bibitem{23} V. Peano and M. Thorwart, Phys. Rev. B {\bf 70}, 235401 (2004).  D. Vitali, S. Gigan, A. Ferreira, H. R. B\"ohm, P. Tombesi, A. ¨
Guerreiro, V. Vedral, A. Zeilinger, and M. Aspelmeyer, Phys.
Rev. Lett. 98, 030405 (2007). M. Karuza {\it et al}., Phys. Rev. A {\bf 88}, 013804 (2013).

\bibitem{24} T. Rocheleau, T. Ndukum, C. Macklin, J. B. Hertzberg, A. A. Clerk, and K. C. Schwab, Nature (London) {\bf 463}, 72 (2010).

\bibitem{25} K. C. Schwab and M. L. Roukes, Phys. Today {\bf 58} (7), 36 (2005).

\bibitem{26} S. A. McGee, D. Meiser, C. A. Regal, K. W. Lehnert, and M. J. Holland, Phys. Rev. A {\bf 87}, 053818 (2013).

\bibitem{27} M. J. Akram and F. Saif, arXiv:1411.0711 [quant-ph] (2014).

\bibitem{28} F. Bariani, J. Otterbach, H. Tan, and P. Meystre, Phys. Rev. A {\bf 89}, 011801(R) (2014).

\bibitem{29} K. Hammerer {\it et al.,} Phys. Rev. Lett. {\bf 103}, 063005 (2009).

\bibitem{30} J. Restrepo, C. Ciuti, and I. Favero, Phys. Rev. Lett. \textbf{112}, 013601 (2014).

\bibitem{31} F. Saif, Phys. Rep. {\bf 419}, 207 (2005); {\bf 425}, 369 (2006).

\bibitem{32} A. Carmele, B. Vogell, K. Stannigel, and P. Zoller, New J. Phys. {\bf 16}, 063042 (2014).

\bibitem{33} F. Brennecke, S. Ritter, T. Donner, and T. Esslinger, Science {\bf 322}, 235 (2008).

\bibitem{34} P. Kanamoto and P. Meystre, Phys. Rev. Lett. {\bf 104}, 063601 (2010).

\bibitem{35} S. Yang, M. Al-Amri, and M. S. Zubairy, J. Phys. B: At. Mol. Opt. Phys. {\bf 47}, 135503 (2014).

\bibitem{36} F. Bariani, S. Singh, L. F. Buchmann, M. Vengalattore,
and P. Meystre, Phys. Rev. A {\bf 90}, 033838 (2014).

\bibitem{37} R. J. Schoelkopf and S. M. Girvin, Nature (London) {\bf 451}, 664 (2008).

\bibitem{38} H. J. Kimble, Nature (London) {\bf 453}, 1023 (2008).

\bibitem{39} J. I. Cirac and P. Zoller, Nature (London) {\bf 404}, 579 (2000).

\bibitem{40} T. D. Ladd et al., Nature (London) {\bf 464}, 45 (2010).

\bibitem{Tian} L. Tian, Ann. Phys. (Berlin) {\bf 527}, 1 (2015).
\bibitem{46} L. Tian and P. Zoller, Phys. Rev. Lett. {\bf 93}, 266403 (2004).
\bibitem{47} W. K. Hensinger, D. W. Utami, H.-S. Goan, K. Schwab, C. Monroe, and G. J. Milburn, Phys. Rev. A {\bf 72}, 041405(R) (2005).
\bibitem{52} Y. D. Wang and A. A. Clerk, Phys. Rev. Lett. {\bf 108}, 153603 (2012). L. Tian, Phys. Rev. Lett. {\bf 108}, 153604 (2012).
\bibitem{522} S. D. Bennett, J. Maassen, and A. A. Clerk, Phys. Rev. Lett. {\bf 105}, 217206 (2010); {\bf 106}, 199902(E) (2011).

\bibitem{44} D. J. Wineland {\it et al}., J. Res. Natl. Inst. Stand. Technol. {\bf 103}, 259 (1998).
\bibitem{45} D. Leibfried, R. Blatt, C. Monroe, and D. Wineland, Rev. Mod. Phys. {\bf 75}, 281 (2003).
\bibitem{Bachtold} P. Weber, J. G\"uttinger, I. Tsioutsios, D. E. Chang, and A. Bachtold, Nano Lett. {\bf 14}, 2854 (2014).

\bibitem{53} J. D. Teufel, J. W. Harlow, C. A. Regal, and K. W. Lehnert,
Phys. Rev. Lett. {\bf 101}, 197203 (2008).

\bibitem{54} A. Schliesser, O. Arcizet, R. Rivi\'ere, G. Anetsberger,
and T. J. Kippenberg, Nature Phys. {\bf 5}, 509 (2009).

\bibitem{55} Y. S. Park and H. Wang. Nature Phys. {\bf 5}, 489 (2009).

\bibitem{56} D. E. Chang, V. Vuleti\'c, and M. D. Lukin Nature Photon. {\bf 8}, 685 (2014). J. D. Thompson {\it et al.,} Nature (London) {\bf 452}, 72 (2008).

\bibitem{57} S. Gr\"oblacher, K. Hammerer, M. R. Vanner, and M.
Aspelmeyer, Nature (London) {\bf 460}, 724 (2009).

\bibitem{Bennett} S. D. Bennett, L. Cockins, Y. Miyahara, P. Gr\"utter, and A. A. Clerk, Phys. Rev. Lett. {\bf 104}, 017203 (2010).

\bibitem{58} M. Asjad, and F. Saif, Optik {\bf 125}, 5455 (2014). S. Huang and G. S. Agarwal, Phys. Rev. A {\bf 81}, 033830 (2010).

\bibitem{59} T. Faust, J. Rieger, M. J. Seitner, J. P. Kotthaus, and E.
M. Weig, Nature Phys. {\bf 9}, 485 (2013). F. Massel {\it et al.,} Nat. Commun. {3}, 987 (2012).

\bibitem{48} K. H. Lee, T. G. McRae, G. I. Harris, J. Knittel, and W. P. Bowen, Phys. Rev. Lett. {\bf 104}, 123604 (2010).
\bibitem{50} S. A. McGee, D. Meiser, C. A. Regal, K. W. Lehnert, and M. J. Holland, Phys. Rev. A {\bf 87}, 053818 (2013).

\bibitem{51} Li, H. K., Ren, X. X., Liu, Y. C. \& Xiao Y. F. Photon-photon interactions in a largely detuned optomechanical cavity. {\it Phys. Rev. A} {\bf 88}, 053850 (2013).

\bibitem{Wang} Y. D. Wang and A. A. Clerk, New J. Phys. {\bf 14}, 105010 (2012).

\bibitem{Tsang} M. Tsang, Phys. Rev. A {\bf 81}, 063837 (2010). M. Tsang, Phys. Rev. A {\bf 84}, 043845 (2011).

\bibitem{Terry} T. G. McRae, K. H. Lee, G. I. Harris, J. Knittel, and W. P. Bowen, Phys. Rev. A {\bf 82}, 023825 (2010).

\bibitem{Regal} C. A. Regal and K. W. Lehnert, J. Phys. Conf. Ser. {\bf 264}, 012025 (2011).

\bibitem{Clerk} A. A. Clerk, X. Waintal, and P. W. Brouwer, Phys. Rev. Lett. {\bf 86}, 4636 (2001).

\bibitem{60} P.-C. Ma, J.-Q. Zhang, Y. Xiao, M. Feng, and Z.-M. Zhang, Phys. Rev. A {\bf 90}, 043825 (2014).

\bibitem{61} H. Wang, X. Gu, Y. X. Liu, A. Miranowicz, and F. Nori, Phys. Rev. A {\bf 90}, 023817 (2014).

\bibitem{63} D. Tarhan, Act. Phys. Pol. A {\bf 124}, 46 (2013). D. Tarhan, S. Huang, and O. E. Mustecaplioglu, Phys. Rev. A {\bf87}, 013824 (2013).

\bibitem{64} B. Chen, C. Jiang, and K. D. Zhu, Phys. Rev. A {\bf 83}, 055803 
(2011).

\bibitem{65} X-G Zhan {\it et al.,} J. Phys. B: At. Mol. Opt. Phys. {\bf 46}, 025501 (2013).

\bibitem{66} C. Jiang, H. Liu, Y. Cui, X. Li, G. Chen, and B. Chen, Opt. Express {\bf 21}, 12165 (2013).

\bibitem{67} G. S. Agarwal and S. Huang, Phys. Rev. A 81, 041803(R) (2010).

\bibitem{68} L. Wang,	Y. Gu, H. Chen,	J. Y. Zhang,	Y. Cui,	B. D. Gerardot, and Q. Gong, Sci. Rep. {\bf 3}, 2879 (2013).

\bibitem{62} S. Huang, J. Phys. B: At. Mol. Opt. Phys. {\bf 47}, 055504 (2014).

\bibitem{69} G. S. Agarwal, T. N. Dey, and S. Menon, Phys. Rev. A {\bf 64}, 053809 (2001).

\bibitem{70} M. Sahrai, H. Tajalli, K. T. Kapale, and M. S. Zubairy, Phys. Rev. A {\bf 70}, 023813 (2004).

\end{thebibliography}
\end{document}